\documentclass[11pt,A4paper]{article}

\usepackage{units}

\usepackage{natbib}
\usepackage{graphicx}
\usepackage{amsmath}
\usepackage{wrapfig}
\usepackage{amssymb}
\usepackage{afterpage}
\usepackage{graphics}
\usepackage{float}
\usepackage{lineno}
\usepackage{rotating}
\usepackage{xspace}
\usepackage{dcolumn} 
\usepackage{bm} 

\usepackage[left=0.7in,top=1.0in,bottom=1.0in,right=0.7in]{geometry}

\title{\textbf{A JVLA 10~degree$^2$ deep survey}}
\date{} 

\newcommand{\be}{\begin{equation}}
\newcommand{\ee}{\end{equation}}

\def\gtsim{\mathrel{\raise0.35ex\hbox{$\scriptstyle >$}\kern-0.6em
\lower0.40ex\hbox{{$\scriptstyle \sim$}}}}
\def\ltsim{\mathrel{\raise0.35ex\hbox{$\scriptstyle <$}\kern-0.6em
\lower0.40ex\hbox{{$\scriptstyle \sim$}}}}
\def\spose#1{\hbox to 0pt{#1\hss}}
\def\lta{\mathrel{\spose{\lower 3pt\hbox{$\mathchar"218$}}
     \raise 2.0pt\hbox{$\mathchar"13C$}}}
\def\gta{\mathrel{\spose{\lower 3pt\hbox{$\mathchar"218$}}
     \raise 2.0pt\hbox{$\mathchar"13E$}}}

\begin{document}

\vspace{-3.5cm}
\maketitle
\renewcommand*{\thefootnote}{\fnsymbol{footnote}}
\vspace{-2.0cm}
\begin{center}
Matt Jarvis$^{1,2}$\footnote{Contact author: matt.jarvis@astro.ox.ac.uk}, Sanjay Bhatnagar$^3$, Marcus Br\"uggen$^4$, Chiara Ferrari$^5$, Ian Heywood$^{6,7}$,
Martin Hardcastle$^8$, Eric Murphy$^9$, Russ Taylor$^{2,10,11}$, Oleg Smirnov$^{7,12}$, Chris Simpson$^{13}$, Vernesa Smol\v{c}i\'c$^{14}$,
Jeroen Stil$^{11}$, Kurt van der Heyden$^{10}$
\end{center}

\renewcommand*{\thefootnote}{\arabic{footnote}}
\setcounter{footnote}{0}
\normalsize
\vspace{-0.50cm}
\begin{abstract}
\bf{One of the fundamental challenges for astrophysics in the 21st century is finding a way to untangle the physical processes that govern galaxy formation and evolution.  
Given the importance and scope of this problem, 
the multi-wavelength astronomical community has used the past decade to build up a wealth of information over specific extragalactic deep fields to address key questions in galaxy formation and evolution. These fields generally cover at least 10~square degrees to facilitate the investigation of the rarest, typically most massive, galaxies and AGN. Furthermore,  such areal coverage allows the environments to be fully accounted for, thereby linking the single halo to the two-halo terms in the halo occupation distribution. Surveys at radio wavelengths have begun to lag behind those at other wavelengths, especially in this medium-deep survey tier.  However, the survey speed offered by the JVLA means that we can now reach a point where we can begin to obtain commensurate data at radio wavelengths to those which already exists from the X-ray through to the far-infrared over $\sim10$ square degrees. 

We therefore present the case for a 10 square degree survey to 1.5$\mu$Jy at L-band in A or B Array, requiring $\sim 4000$  hours to provide census of star-formation and AGN-accretion activity in the Universe. For example, the observations will allow galaxies forming stars at 10~M$_{\odot}$~yr$^{-1}$  to be detected out to $z\sim 1$ and luminous infrared galaxies (1000~M$_{\odot}$~yr$^{-1}$) to be found out to $z \sim 6$. Furthermore, the survey area ensures that we will have enough cosmic volume to find these rare sources at all epochs. The bandwidth will allow us to determine the polarisation properties galaxies in the high-redshift Universe as a function of stellar mass, morphology and redshift. Conducting the a survey at L-band ensures the highest sensitivity for detecting extragalactic radio sources, even with the reduced effective bandwidth of 40 per cent due to RFI, compared to S-band, over such an area. However, an additional crucial aspect of carrying out this survey at L-band is that it could be carried out jointly with a deep and wide H{\sc i}-survey over fields which have similar amounts of spectroscopic and imaging data as currently available to CHILES, and given the number of repeat observations, the survey will also be ideal for extending the search for faint radio transients.

Given the existing and planned multi-wavelength efforts by the astronomical community, the survey fields most appropriate for such a JVLA effort are the XMMLSS field and the Extended Chandra Deep-Field South, which when combined with the COSMOS/CHILES survey will provide around 10 square degrees of radio continuum, polarisation and spectral-line data over the best studied regions of the extragalactic sky.
}
\end{abstract}


\section{Science Justification}\label{sec:science}

\noindent
During the first billion years of cosmic time, primordial galaxies emerged from a near-featureless early universe and began assembling into the chemically complex and morphological diverse population of galaxies that populate the present-day cosmos. The nature of this process, how stars and galaxies have emerged and evolved from the Big Bang to today, is one of the most compelling questions in all of science,  resulting in a community-wide, multi-wavelength effort to obtain a wealth of information over carefully selected extragalactic deep fields to specifically address key questions in galaxy formation and evolution.   The one waveband that is currently lagging is the radio, however, with the dramatically increased capabilities of the JVLA, we can finally begin to obtain radio data at commensurate depths with other multi-wavelength efforts, spanning from the X-ray through the far-infrared over $\sim10$ square degrees. 
To this end, we suggest that the VLASS has a deep continuum survey component that will provide the necessary depth to investigate the evolution of the radio source population from the Epoch of Reionization to the present day. The depth of the radio observations suggested is designed to fully sample the luminosity function of both AGN-powered radio sources and the star-forming galaxy population, from redshifts $z=0 \rightarrow 6$. Such a survey will also be sufficiently wide in area that it will enable us to overcome cosmic variance issues and be able to assess the evolutionary status of galaxies as a function of their environment. We outline the key science cases for the continuum survey below, critically select the best survey fields for the VLASS in Section~\ref{sec:multilambda}, and provide detailed description of the survey strategy in Section~\ref{sec:strategy}.


\subsection{Cosmic star-formation history and galaxy evolution}

\noindent
With a deep continuum survey with the JVLA, we aim to understand the basic features of galaxy formation and evolution, the volume-averaged star formation rate as a function of epoch, its distribution function within the galaxy population, and its variation with environment. Surveys of the star-formation rate (SFR) as a function of epoch suggest that the star-formation rate density
rises as $\sim (1 + z)^4$ out to at least $z \sim 1$ \citep[e.g.][]{1996ApJ...460L...1L,2013ApJ...770...57B} and then flattens, with the bulk of stars seen in galaxies today having been formed between $z \sim 1-3$. Determining the precise redshift where the star-formation rate peaked is more difficult, with different star-formation indicators giving widely different measures of the integrated star-formation rate density \citep[see e.g.][]{2006ApJ...651..142H}. These problems are exacerbated by the effects of cosmic variance in the current samples (multi-wavelength surveys such as COSMOS and GOODS typically cover only modest-sized areas, $\ltsim 1$~degree$^2$, corresponding to just $\sim 30$~Mpc at $z > 1$), as well as small sample sizes.

\noindent
Below $S_{\rm 1.4GHz} \sim 200\mu$Jy, SFGs begin to dominate the radio source population \citep[e.g.][]{1985ApJ...289..494W, 2008MNRAS.388.1335W}. The radio continuum emission of these galaxies offers a star-formation indicator \citep[e.g.][]{Yun01} that can be studied across the history of the Universe, and in particular right through the peak star-formation epoch at $1 \ltsim z \ltsim 3$. 
A JVLA continuum survey that covers a relatively large sky area but which is also sensitive enough that estimates of the integrated star-formation rate do not require large extrapolations for faint sources is required. An example of a radio continuum survey that would lead this field for years to come would be to reach star-formation rate limits of $\sim 10$~M$_{\odot}$~yr$^{-1}$ and $\sim 30$~M$_{\odot}$~yr$^{-1}$, at $z \sim 1$ and $z\sim 2$ respectively, which is comparable to the star-formation rate sensitivity of deep far-infrared surveys conducted with the SPIRE instrument on {\em Herschel}. 
Crucially, radio continuum is also immune to dust extinction, and will therefore be able to identify heavily dust-obscured systems, often missed in optically-selected samples. This will be particularly valuable in combination with {\em Herschel}, where the high-spatial-resolution radio data will be used to de-confuse the low-spatial-resolution images and investigate the evolution of the far-infrared--radio relation \citep[e.g.][]{Appleton04,2009ApJ...706..482M,Sargent10,2010A&A...518L..31I,2010MNRAS.409...92J}, and thus how radio emission relates to star-formation rate in galaxies at the highest redshifts.
It is not only the global average star formation rate that is important for our understanding of galaxy formation and evolution, but more crucially the nature and distribution of the star-forming galaxies at high redshifts:

\noindent
{\it How does star-formation proceed as a function of galaxy mass?:} it is well established that in the local Universe the stellar populations of the most massive galaxies formed earlier than those of less massive galaxies \citep[`down-sizing'; e.g.][]{1996AJ....112..839C} and so massive galaxies must form stars rapidly at an early epoch, and then have their star formation truncated, but how and exactly when did this occur?

\noindent
{\it What is the role of galaxy environment?:} in the local Universe, star formation is suppressed in dense environments \citep[e.g.][]{2002MNRAS.334..673L}, an effect which diminishes with increasing redshift, with hints that it disappears altogether at $z\sim2$ \citep[e.g.][]{2013MNRAS.434..423K, 2014MNRAS.437..458Z}. But where precisely, in terms of epoch and environment, does this environmental influence begin to become important? To what extent is the build-up of galaxies into groups and clusters responsible for the sharp decline in the global average star formation rate below $z = 1$. Is this indeed an environmental effect or host galaxy mass dependent effect \citep[e.g.][]{2010ApJ...721..193P, 2012ApJ...757....4P}?

\noindent
{\it How does star formation relate to the growth of supermassive black holes, and AGN feedback?:} it is widely believed that AGN activity (particularly radio-loud AGN) may be responsible for switching off star-formation in massive galaxies, but a direct observational link between AGN activity and star-formation at high redshifts remains elusive. Indeed, recent studies from both a theoretical \citep{2013ApJ...772..112S} and observational \citep{2012MNRAS.427.2401K} perspective have shown that powerful radio-loud AGN may actually provide a positive form of feedback. On the other hand, there is little evidence for any type of feedback from radio-quiet objects based on the latest studies using {\em Herschel} \citep[e.g.][]{2011MNRAS.416...13B,2013A&A...560A..72R}
Given that different forms of AGN feedback are invoked in semi-analytic models \citep[e.g.][]{2006MNRAS.365...11C, 2006MNRAS.370..645B,2012MNRAS.420L...8H} of galaxy formation we are required to understand such processes if we are ever to understand the evolution of galaxies.

\noindent
A deep JVLA continuum survey is required to provide definitive answers to all of these questions, free from cosmic variance \citep[e.g.][]{2013MNRAS.432.2625H} or dust extinction biases. A survey that will detect upward of 10,000 star-forming galaxies (to a $5\sigma$ limit) in each of the redshift ranges $1 < z < 1.5$, $1.5 < z < 2$, $2 < z < 3$ and around 5000 sources at  $z>3$ would allow for detailed analyses of different sub-populations (e.g. dividing into $10-100$ bins in planes such as SFR vs stellar mass will allow $\gtsim$100 galaxies per bin). The exquisite-quality multi-wavelength datasets available in key extragalactic fields will provide spectroscopic redshifts where dedicated surveys have been or are being undertaken and excellent photometric redshifts from the deep optical and infrared photometry. Such data will also provide source properties such as masses and environments, and classifications as AGN or mergers. In Figure~1 we show the expected constraints on the radio-luminosity function for star-forming galaxies at six different epochs for a range of survey areas. One can immediately see the requirement to move to both a deep and reasonably wide survey in order to address the key science of the evolution of star-formation activity in the Universe.

It is also worth emphasising that such a survey conducted over well-studied deep fields also allows the radio luminosity function to be probed to flux-density levels below the nominal survey depth, through traditional stacking \citep[e.g.][]{2011ApJ...730...61K,2014arXiv1401.1648Z} or through new methods \citep[e.g.][]{2014MNRAS.437.2270M,2013arXiv1312.5347R} to retain all information possible and enhance the power of radio continuum surveys in probing the star-formation properties of optically-selected galaxies.

\subsection{The evolution of accretion activity and the exploration of the epoch of reionization} 

\noindent
Active Galactic Nuclei (AGNs) play a major role in the framework of
galaxy formation. The enormous amount of energy they release in the
form of ionising radiation or relativistic jets during their short
lifetime can have a significant effect on their surroundings. The
energetic feedback from AGN appears to be a vital ingredient for
reproducing some observed features, such as the stellar galaxy mass
function \citep{2006MNRAS.365...11C,2006MNRAS.370..645B}, and the black-hole mass
versus bulge mass (or velocity dispersion) correlation \citep{1998AJ....115.2285M,2000ApJ...539L..13G,2000ApJ...539L...9F}. This
suggests that AGN and star formation activity may have been
concurrent, even though the origins of the above correlations are not
entirely clear and the subject of active debate. Indeed, the peak
of QSO activity appears to take place at $z \sim 2$ \citep[e.g.][]{2005A&A...441..417H,2009MNRAS.399.1755C}, i.e. at epochs when
star formation was also at its peak, and observational evidence has been found of the presence of
an embedded AGN in 20-30\% of $z \sim 2$ massive star forming galaxies
in the GOODS fields \citep{2007ApJ...670..173D}. Traditionally the provinces
of separate research fields, it is now becoming clear that AGN and star-formation
activity are intimately related, and the cosmic star formation rate
appears to mirror closely the cosmic accretion rate onto AGN. However,
a complete census of both star formation and AGN activity, especially at
high redshifts, is complicated due to dust extinction and gas
obscuration by circumnuclear material. The optimal combination of 
sensitivity and spatial resolution of JVLA would allow the study of the
entire AGN population from classical radio-loud sources down to the realm of 
radio-quiet AGNs \citep[$P\sim 10^{22-23}$ W Hz$^{-1}$; ][]{2004NewAR..48.1173J,2008MNRAS.388.1335W,2011ApJ...739L..29K,2013ApJ...768...37C}, from $z=0 \rightarrow 6$. This would provide, for the first time, a complete view 
of nuclear activity in galaxies and of its evolution, unbiased by
gas/dust selection effects.


\begin{figure*}
\begin{center}
\resizebox{5.2cm}{!}{\includegraphics{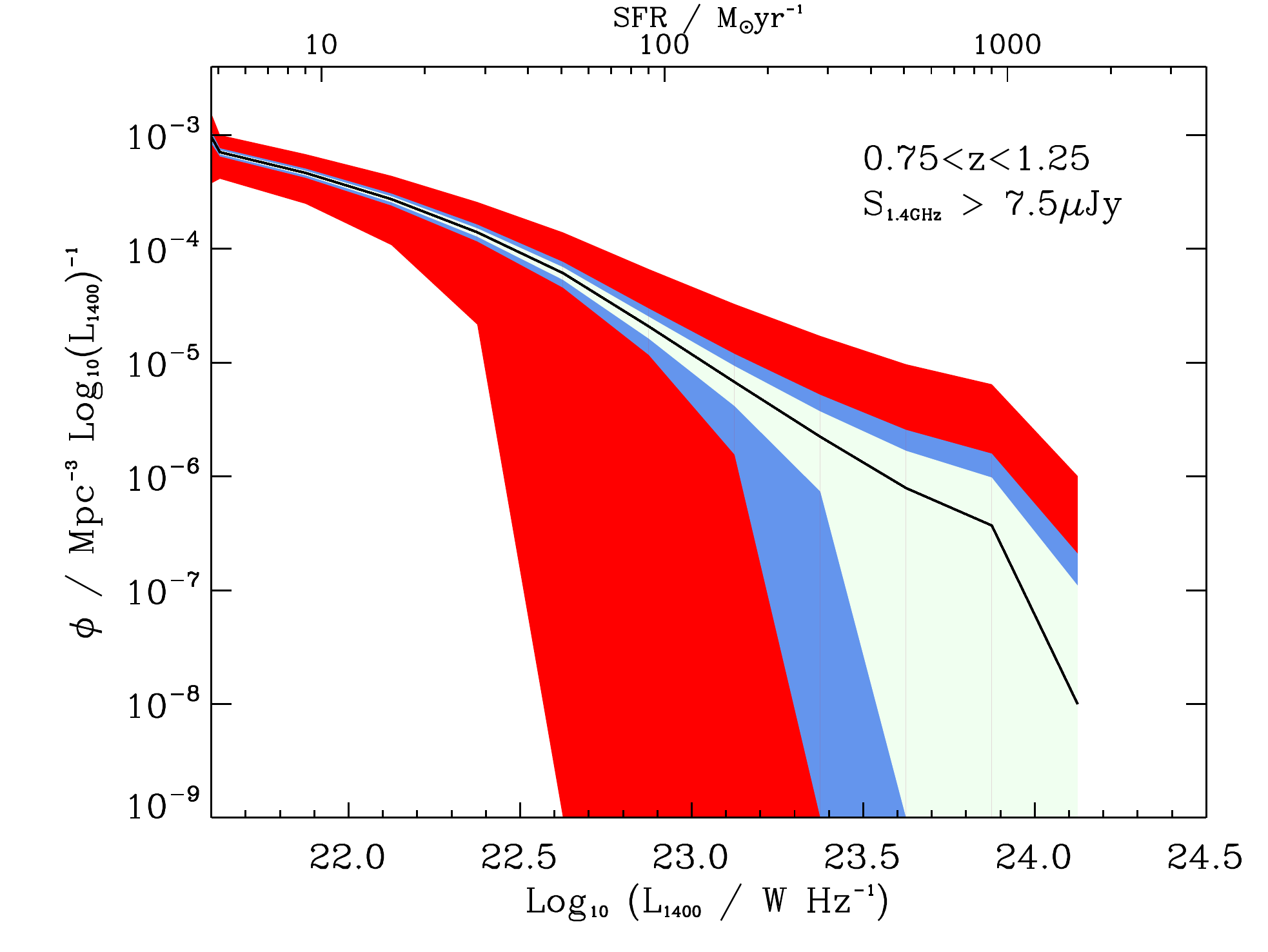}}
\resizebox{5.2cm}{!}{\includegraphics{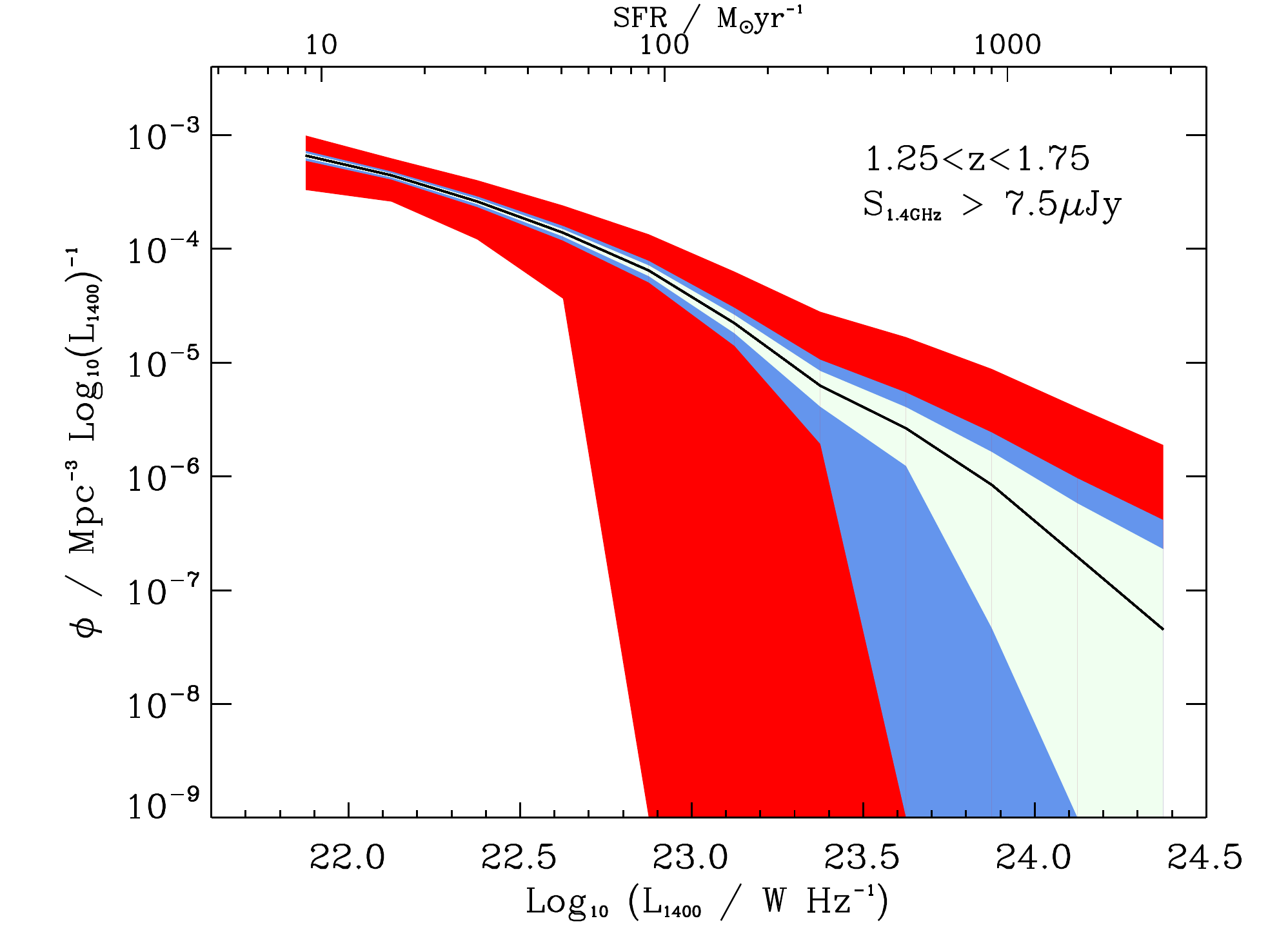}}
\resizebox{5.2cm}{!}{\includegraphics{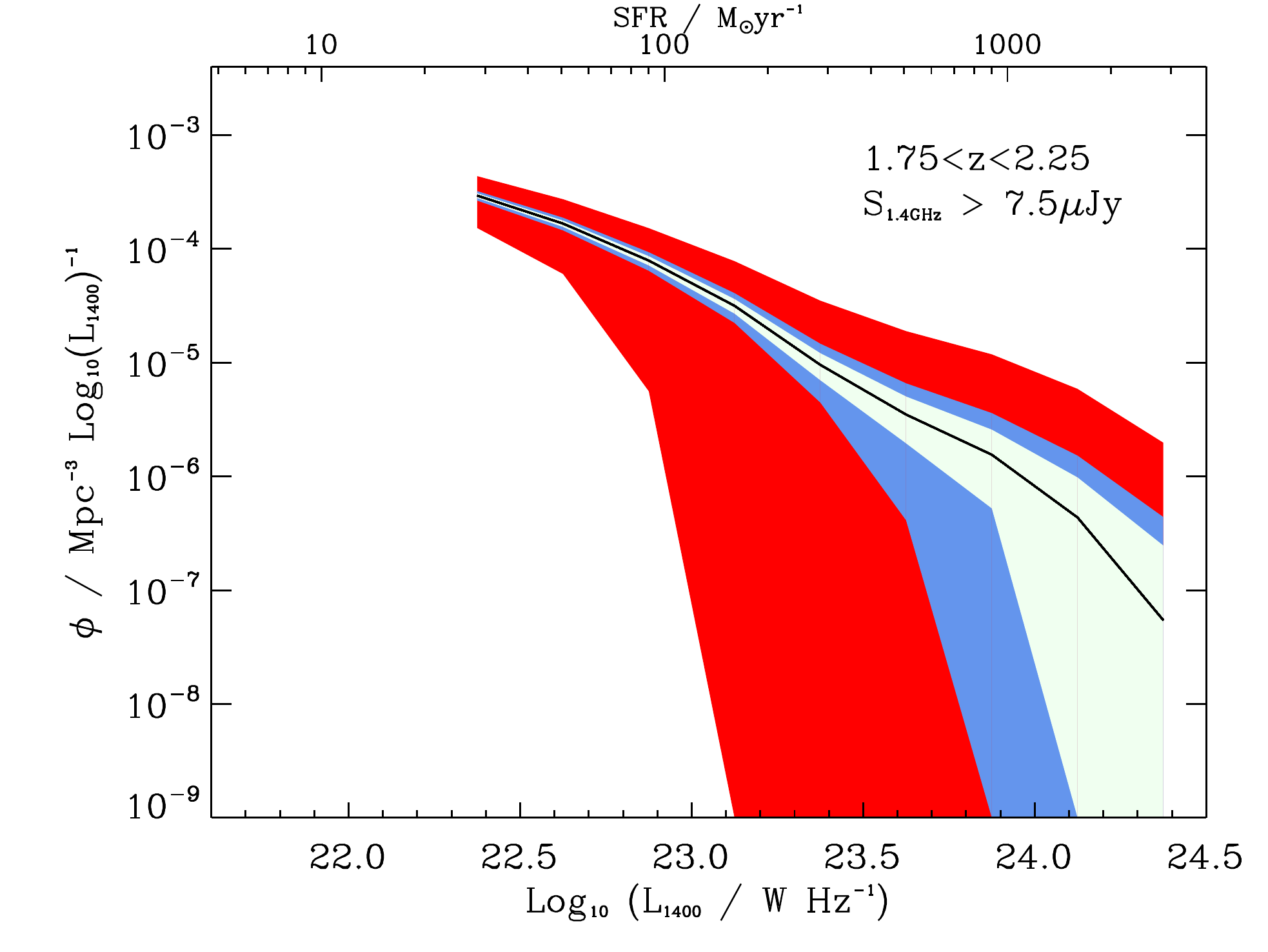}}
\resizebox{5.2cm}{!}{\includegraphics{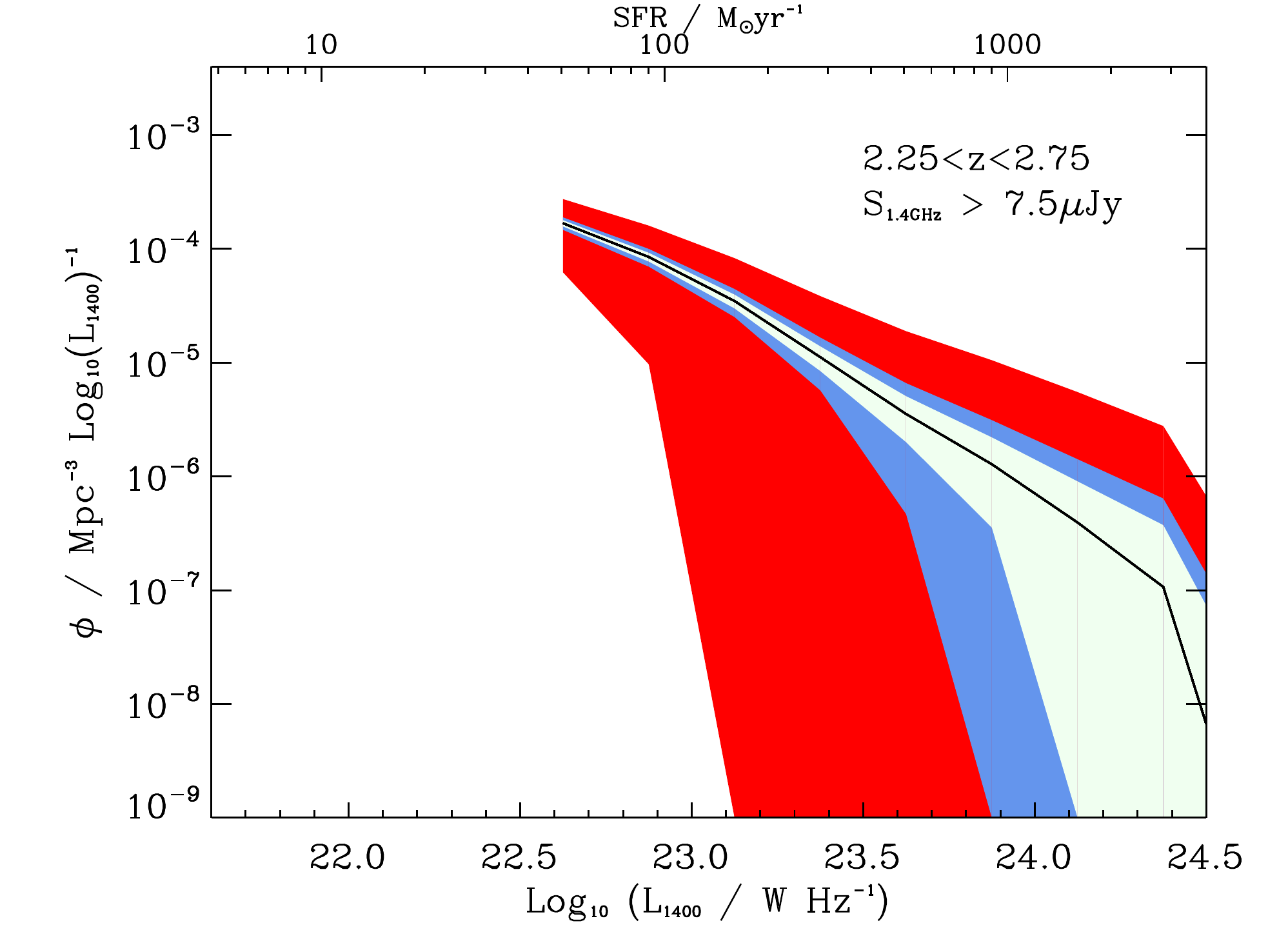}}
\resizebox{5.2cm}{!}{\includegraphics{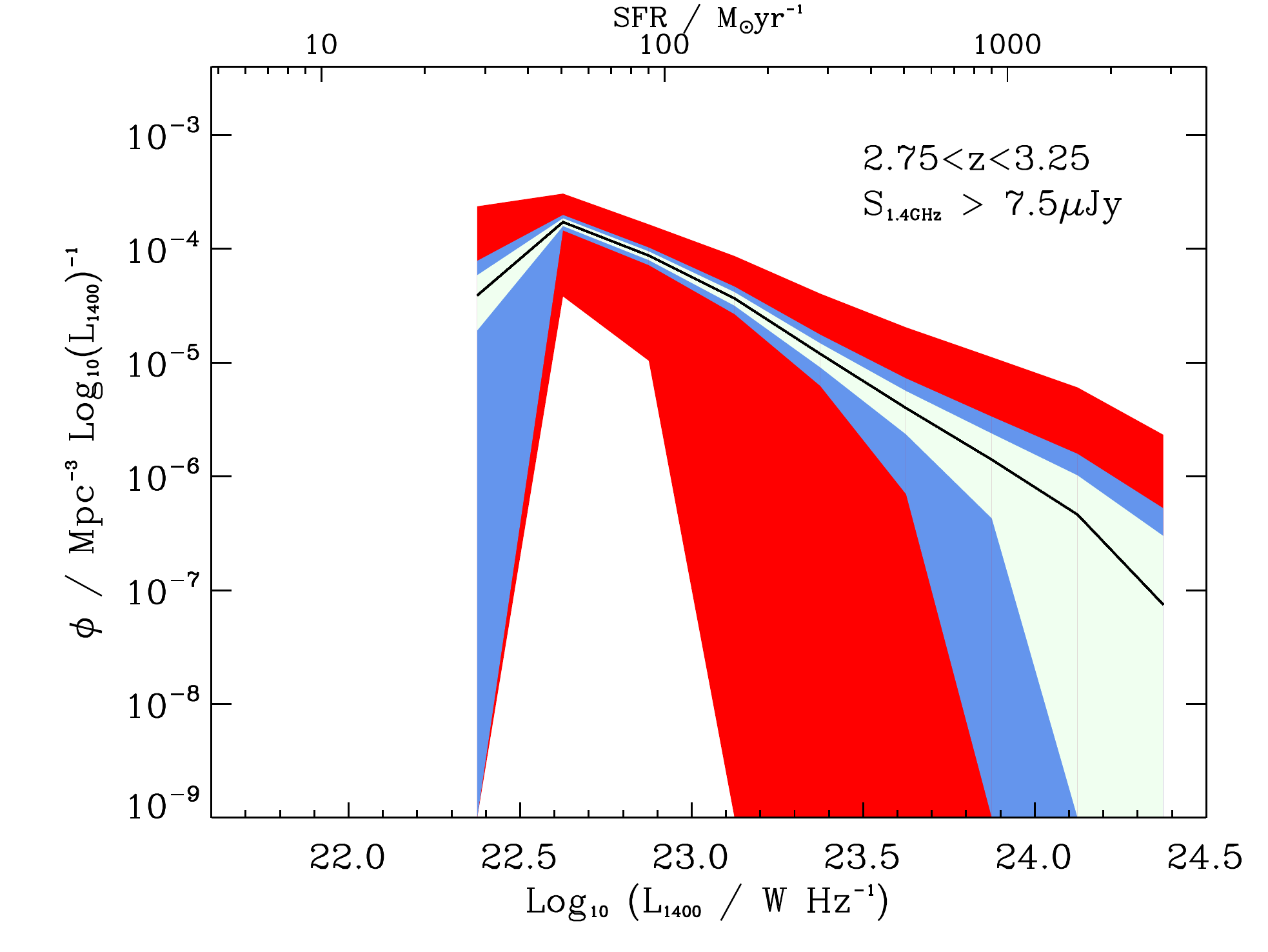}}
\resizebox{5.2cm}{!}{\includegraphics{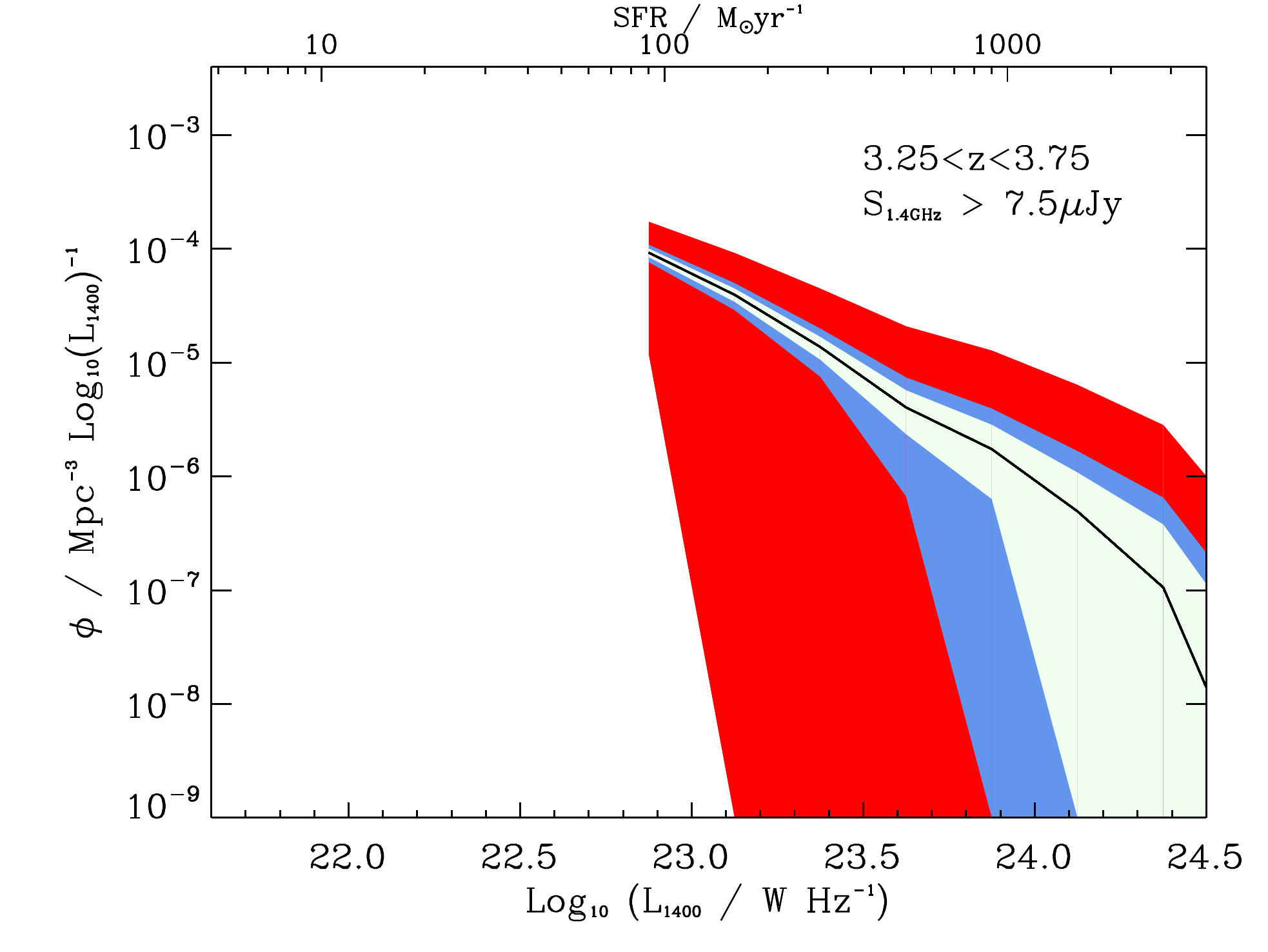}}
\caption{\small\textsf{Predicted luminosity functions in six redshift bins for star-forming galaxies in the deep continuum survey (based on the simulations of \cite{2008MNRAS.388.1335W, 2010MNRAS.405..447W}). The red regions shows the Poisson uncertainty for a 1 degree$^2$ survey, the blue region is for a 5~degree$^2$ survey and the cream regions is for the proposed 10~degree$^2$ survey. The equivalent star-formation rate is given on the upper x-axis.}}
\end{center}
\end{figure*}

\noindent
Indeed, there is now strong evidence that the standard AGN unification paradigm \citep[e.g.][]{1993ARA&A..31..473A,1995PASP..107..803U} for radio-loud AGN does not give a complete picture. For example, observational evidence \citep[e.g.][]{2007MNRAS.376.1849H,2010MNRAS.406.1841H, 2012MNRAS.421.1569B} suggests that many or most low-power ($P < 10^{25}$~W\,Hz$^{−1}$) radio galaxies in the local universe (the numerically dominant population) correspond to a distinct type of AGN. These sources accrete through a radiatively inefficient mode (the so-called `radio mode'), rather than the radiatively efficient accretion mode typical of radio-quiet optically or X-ray selected AGN (`quasar mode'). The role of these two accretion modes appears to be strongly influenced by the environment \citep[e.g.][]{2008A&A...490..893T} while the presence or absence of a radio-loud AGN appears to be a strong function of the stellar mass of the host galaxy \citep[e.g.][]{2005MNRAS.362...25B,2012A&A...541A..62J}. Deep radio surveys covering areas of sky with the best multi-wavelength data (Section~2) will allow us to probe the evolution of these relationships and the accretion mode dichotomy over cosmic time; this is key information for any attempt to incorporate mechanical feedback from radio-loud AGN in models of galaxy, group and cluster formation and evolution. Such studies also drive the need to carry out the radio survey over the best studied fields, and importantly with tracers of the AGN and/or star-formation components in galaxies, such as spectroscopy and full spectral energy distribution modelling \citep[see e.g.][]{2012MNRAS.421.3060S,2013MNRAS.436.1084M,2013MNRAS.433.2647S}. Such a case also pushes for high spatial resolution in order to distinguish jet phenomena from the more abundant emission from star-formation.



\noindent
Furthermore, the details of the mechanism(s) of interaction between radio-loud AGN and their environments, on all scales, remain unclear; such basic questions as whether the most powerful sources are expanding supersonically throughout their lifetimes \citep[e.g.][]{1989ApJ...345L..21B,2000MNRAS.319..562H} or what provides the pressure supporting the lobes of low-power objects \citep[e.g.][]{2008ApJ...686..859B,2008A&A...487..431C} remain unanswered. These questions can only be addressed by the accumulation of large, statistically complete samples of radio sources with good imaging and excellent, homogeneous multi-wavelength data. Information on both large and small-scale radio structure is required. The JVLA deep continuum survey data would not only enable such an investigation for the first time, but also
ensures that the radio luminosity function of all types of AGN is fully sampled at all redshifts (Fig.~2), and covers enough area that the necessary statistical analysis for such work is not severely hampered by small volumes. Furthermore, the vast array of multi-wavelength data will play a crucial role in determining the level of AGN activity in the galaxies, in addition to providing both spectroscopic and photometric redshift information and the immediate environmental density. Such studies are crucial in addressing how galaxies and their supermassive black holes build up together over cosmic time.

\noindent
The key astrophysical problems related to AGN that the proposed survey will address thus include: 

\noindent
1) the relationship between AGN and star-formation activity \citep[e.g.][]{2011MNRAS.416...13B,2013A&A...560A..72R};

\noindent
2) the evolution of low power AGN (including radio-quiet AGN), exploring the so-called ``AGN cosmic downsizing'' scenario, found for X-ray and optically selected AGNs \citep{2005A&A...441..417H, Babic2007, 2009ApJ...696...24S,2011MNRAS.413.1054M,2013MNRAS.436.1084M}; 

\noindent
3) the relative contribution of different accretion regimes (radio vs. quasar modes), its evolution with redshift, and the role played by the environment \citep[e.g.][]{2012MNRAS.421.1569B};

\noindent
4) the relative contribution of radiative versus jet-driven (kinetic) feedback to the global AGN feedback in models of galaxy formation; 

\noindent
5) the mechanisms of that feedback and the evolution in the physical properties of radio-loud AGN with redshift \citep[e.g.][]{2013MNRAS.432.3381M}.

\begin{figure*}
\begin{center}
\resizebox{5.cm}{!}{\includegraphics{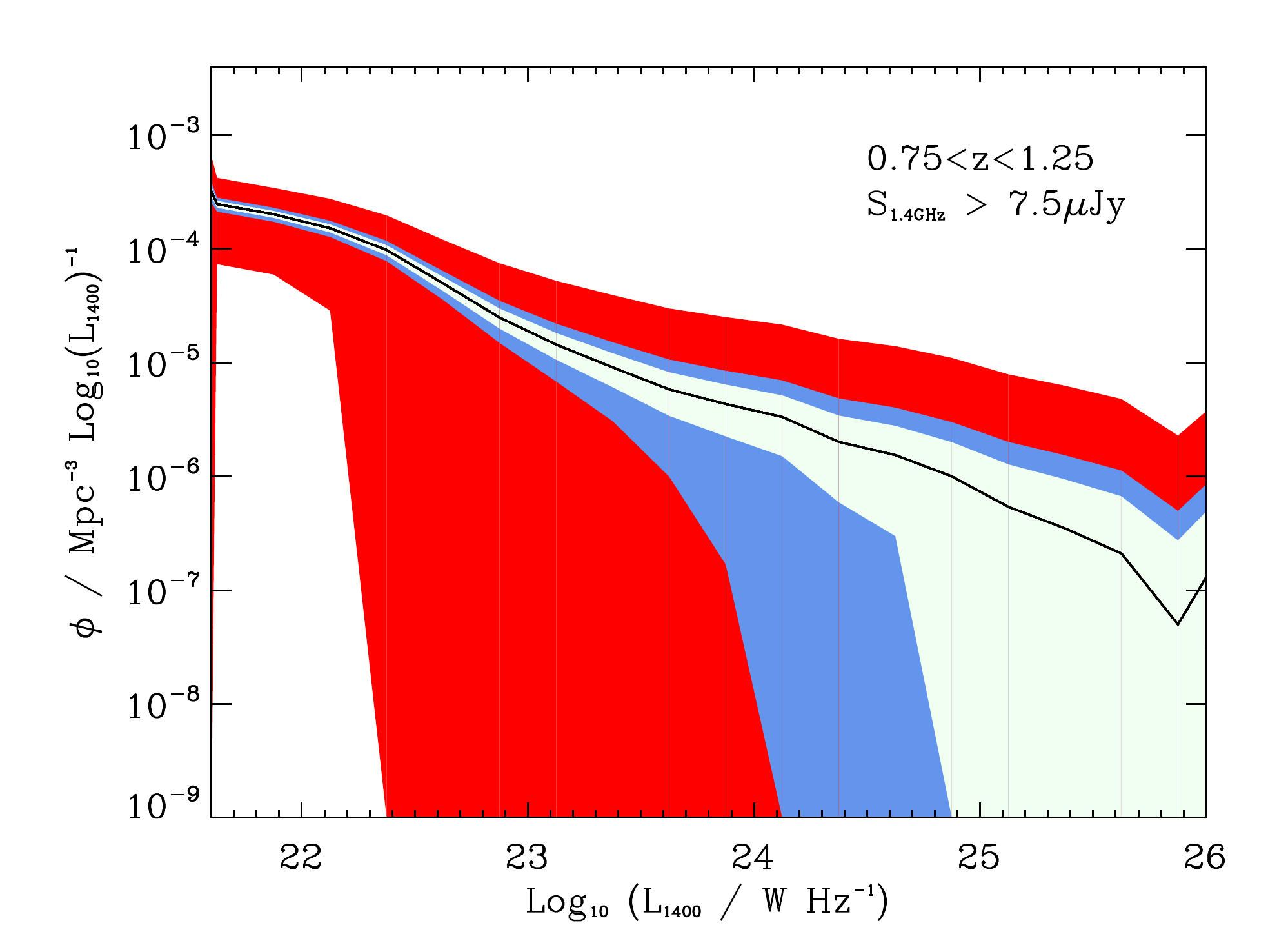}}
\resizebox{5.cm}{!}{\includegraphics{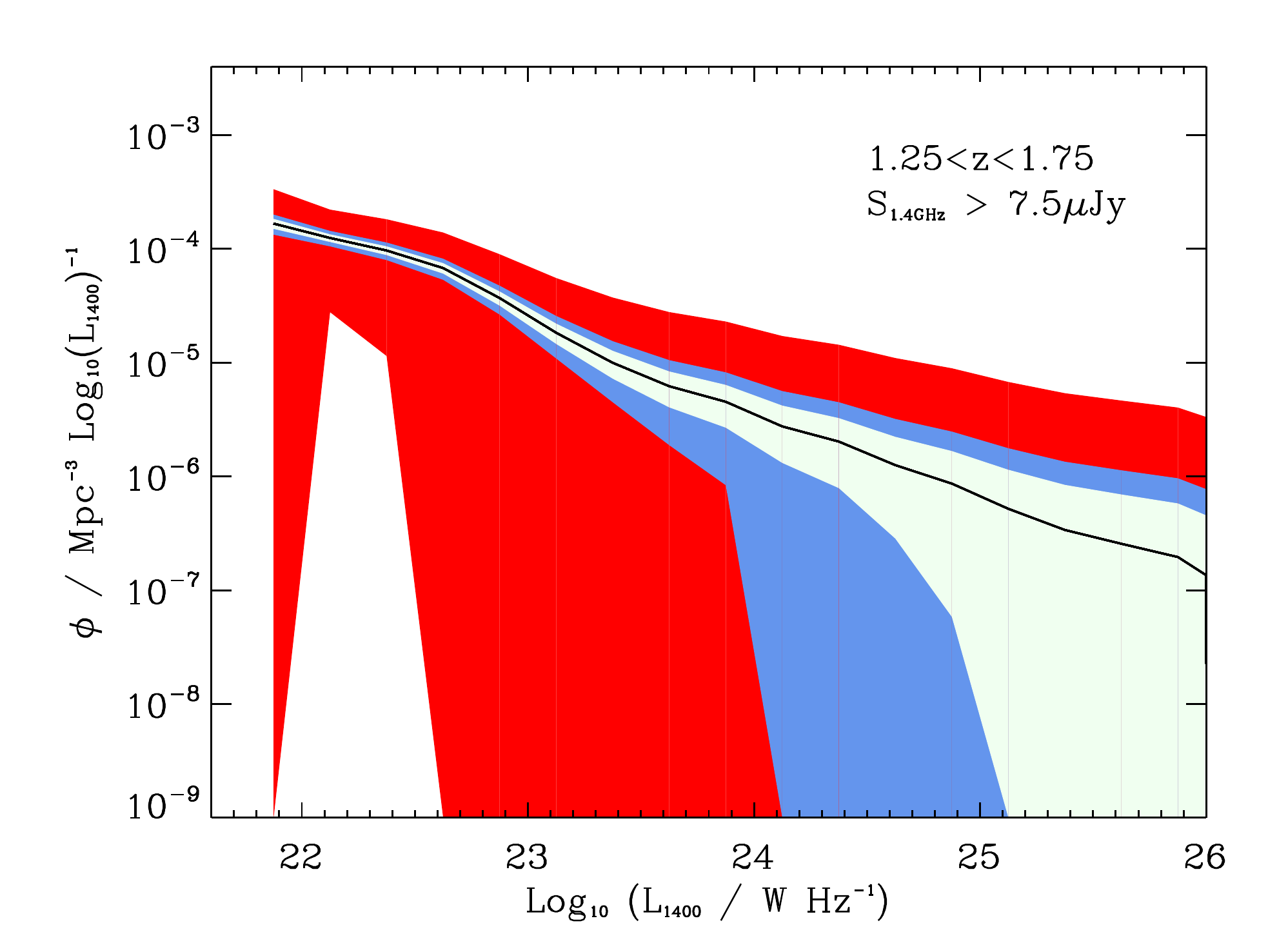}}
\resizebox{5.cm}{!}{\includegraphics{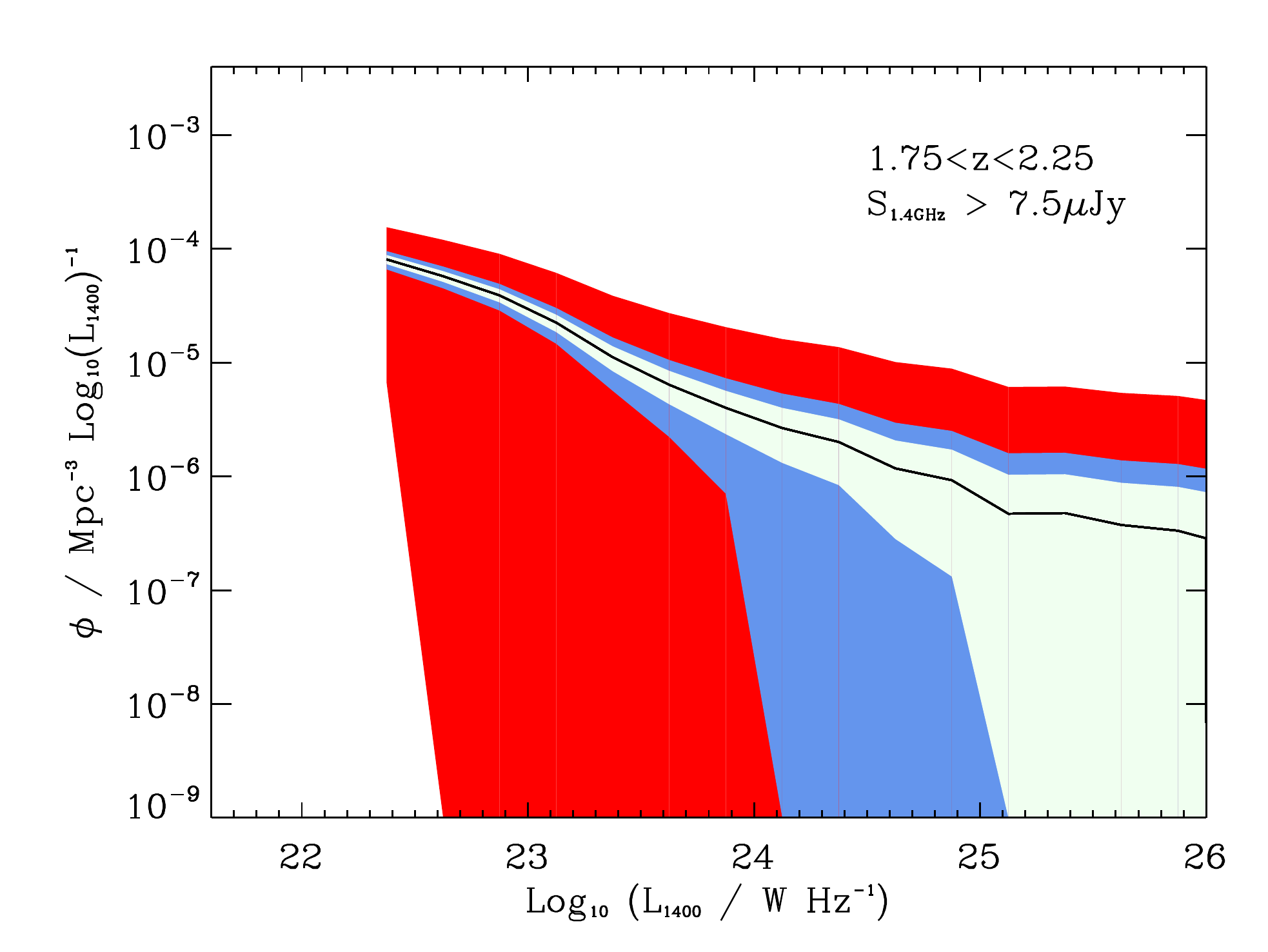}}
\resizebox{5.cm}{!}{\includegraphics{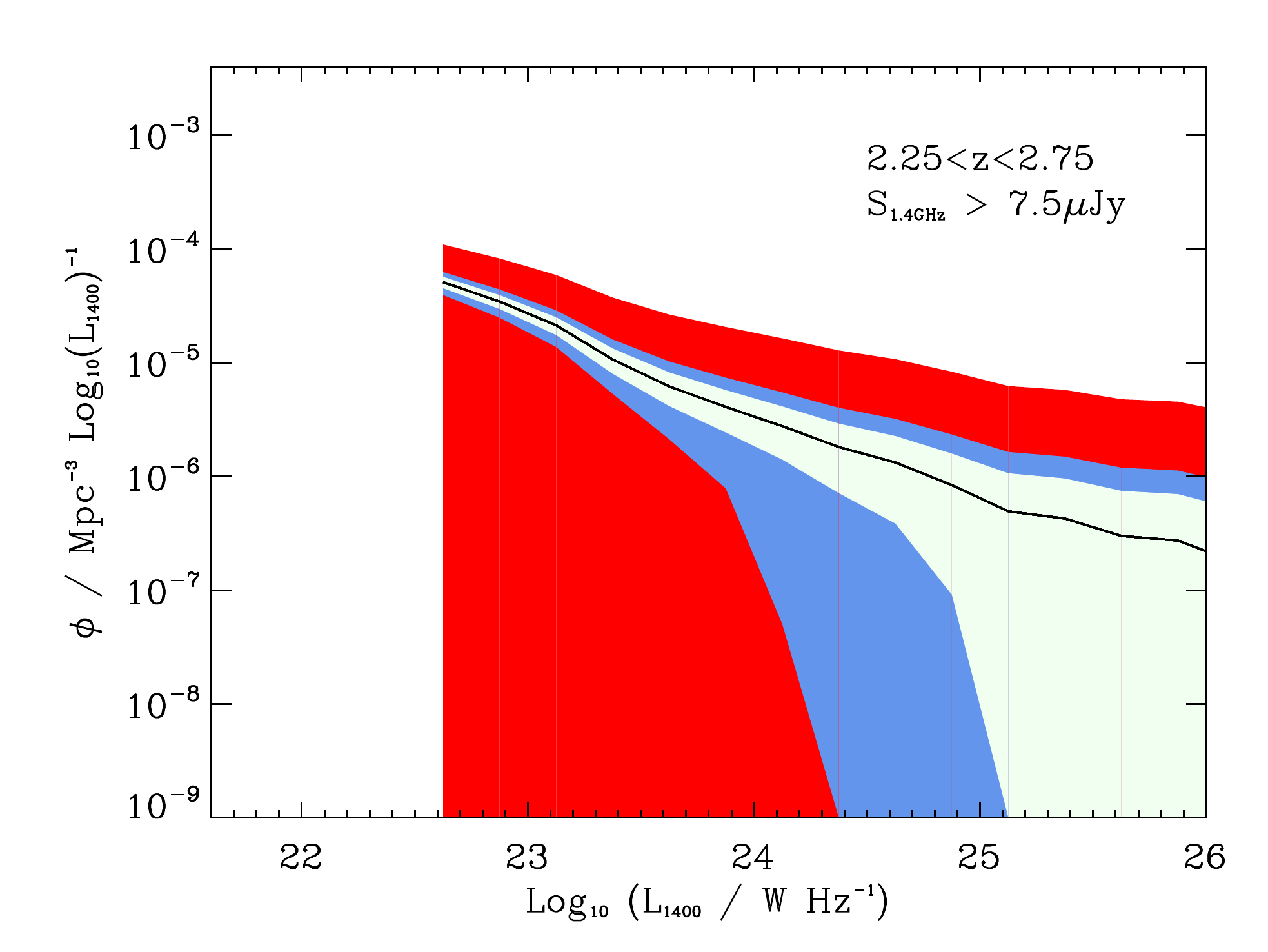}}
\resizebox{5.cm}{!}{\includegraphics{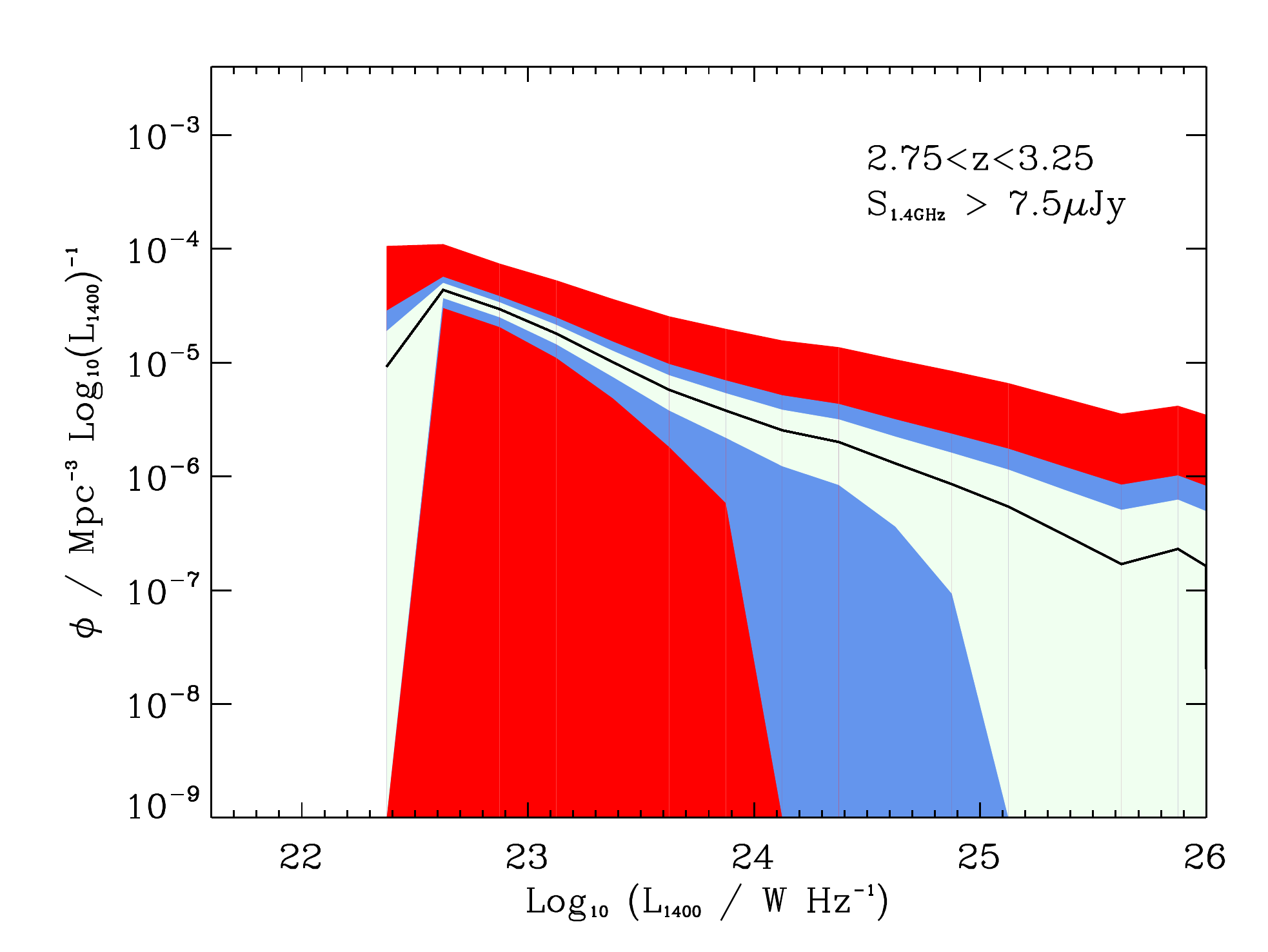}}
\resizebox{5.cm}{!}{\includegraphics{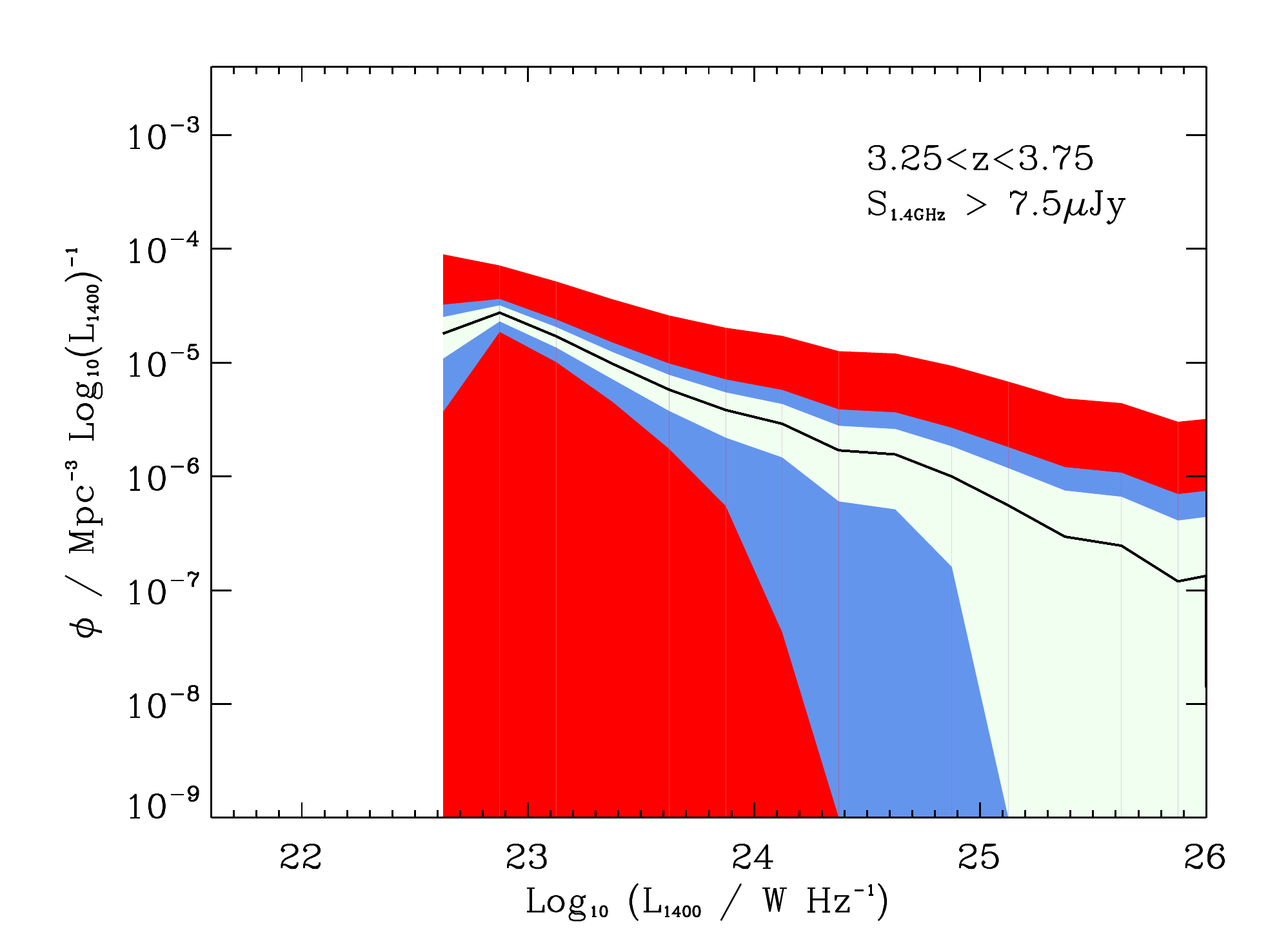}}
\caption{\small\textsf{Predicted luminosity functions in six redshift bins for AGN in the deep continuum survey (based on the simulations by \cite{2008MNRAS.388.1335W, 2010MNRAS.405..447W}). The red regions shows the Poisson uncertainty for a 1 degree$^2$ survey, the blue region is for a 5~degree$^2$ survey and the cream regions is for the proposed 10~degree$^2$ survey. One can see the huge gain in accuracy by mosaicing to 10~degree$^2$ over existing $\sim 1$~degree$^2$ surveys.}}
\end{center}
\end{figure*}

\smallskip
\subsection{Galaxy Clusters}\label{sec:clusters}

The evolution of AGN and star-formation activity can only be understood if the full range in environmental density is probed at all cosmic epochs
\citep[e.g.][]{2008A&A...490..893T,2013MNRAS.430.3086G,Karouzos2014}. Clusters of galaxies represent some of the most extreme environments experienced by galaxies and are therefore ideal laboratories for differentiating the physical processes which can affect and transform the morphologies and star-formation properties of galaxies and their level of AGN activity. Although most work has focused on the suppression of star-formation activity in galaxies as they become part of the cluster population, recent evidence from a small number of radio and mid-infrared studies has demonstrated that there must also be a period of star-formation enhancement to explain the growth of the galactic bulges.  However, the limited size of these studies means that they require confirmation. For example, work combining X-ray, optical, near-infrared, and deep radio observations \citep{2009MNRAS.395...11V,2013ApJ...768....1M} have shown that increased star-formation and/or AGN activity in clusters may be linked to their state of evolution, with more virialized clusters seeming to contain a more quiescent galaxy population than their younger, yet to virialize counterparts. However, such studies are currently very limited in size, e.g. the study of van Breukelen et al. contains just four clusters (or overdensities) at $z \sim 1$ in $\sim 0.5$~degree$^2$. The depth and breadth offered by a 10 square degree deep continuum survey with the JVLA would provide the data to investigate the evolution of the cluster population from $z = 0.5$  to the highest redshifts, when the first clusters are believed to start virializing ($z \sim 2$).

\noindent
To detect typical star forming galaxies at redshifts where there is strongest evidence for environmentally-driven evolution, at $z\sim 0.5-1$, requires $<10$~$\mu$Jy sensitivity to reach star-formation rates of $\sim 10$~M$_{\odot}$~yr$^{-1}$. 
A VLASS JVLA deep continuum survey provides a unique opportunity to study these processes since (a) it is sensitive to the radio continuum emission produced by star-formation and AGN activity; (b) around 10 square degrees are required to serendipitously detect $\sim 100$ of clusters either through radio observations alone or more easily by combining with the deep optical, infrared, SZ and X-ray data over the key extragalactic fields in the $z >0.5$ Universe; and (c) it has the angular resolution required to identify radio emission from individual galaxies within the clusters.

\subsection{Cosmology and Large Scale Structure}

Over the past few years there has been an increasing focus on using radio continuum surveys to address the fundamental issues related to the cosmological model, including determining the equation of state of dark energy and whether we can find evidence for departures from General Relativity on the largest scales \citep[e.g.][]{2012MNRAS.424..801R,2012MNRAS.427.2079C}.
Three key tests where one can use radio continuum sources as cosmological probes are: the Integrated Sachs-Wolfe effect \citep[e.g.][]{2008MNRAS.386.2161R}; the power spectrum of the radio source populations \citep[e.g.][]{2004MNRAS.351..923B}; and the cosmic magnification bias \citep[e.g.][]{2005ApJ...633..589S}. However, one of the key unknowns in our understanding of how well radio sources can help determine the underlying cosmological model is their bias, i.e. how they trace the underlying dark matter density field.

\noindent
It is actually very difficult to determine this quantity directly from radio continuum surveys alone, although some progress has been made by measuring the angular two-point correlation function of radio sources cross-correlated with optical imaging and spectroscopic surveys \citep[e.g.][]{2013MNRAS.429.2183P} and by assuming a redshift distribution \citep[e.g. from the SKADS simulation of][]{2008MNRAS.388.1335W,2010MNRAS.405..447W}. However, such studies are hampered by only the low-redshift sources having reliable optical counterparts, thus limiting the redshift range over which the bias can be measured to $z < 0.5$. Given that the unique niche occupied by radio continuum surveys for determining the cosmological model lie in the fact that their redshift distribution peaks at $1 < z < 2$ (depending on the precise flux-density limit), our lack of knowledge of the bias at $z > 1$ hampers our ability to use these sources as tracers of the Universe on large scales.

\begin{wrapfigure}{l}{9.5cm}
\resizebox{9.3cm}{!}{\includegraphics{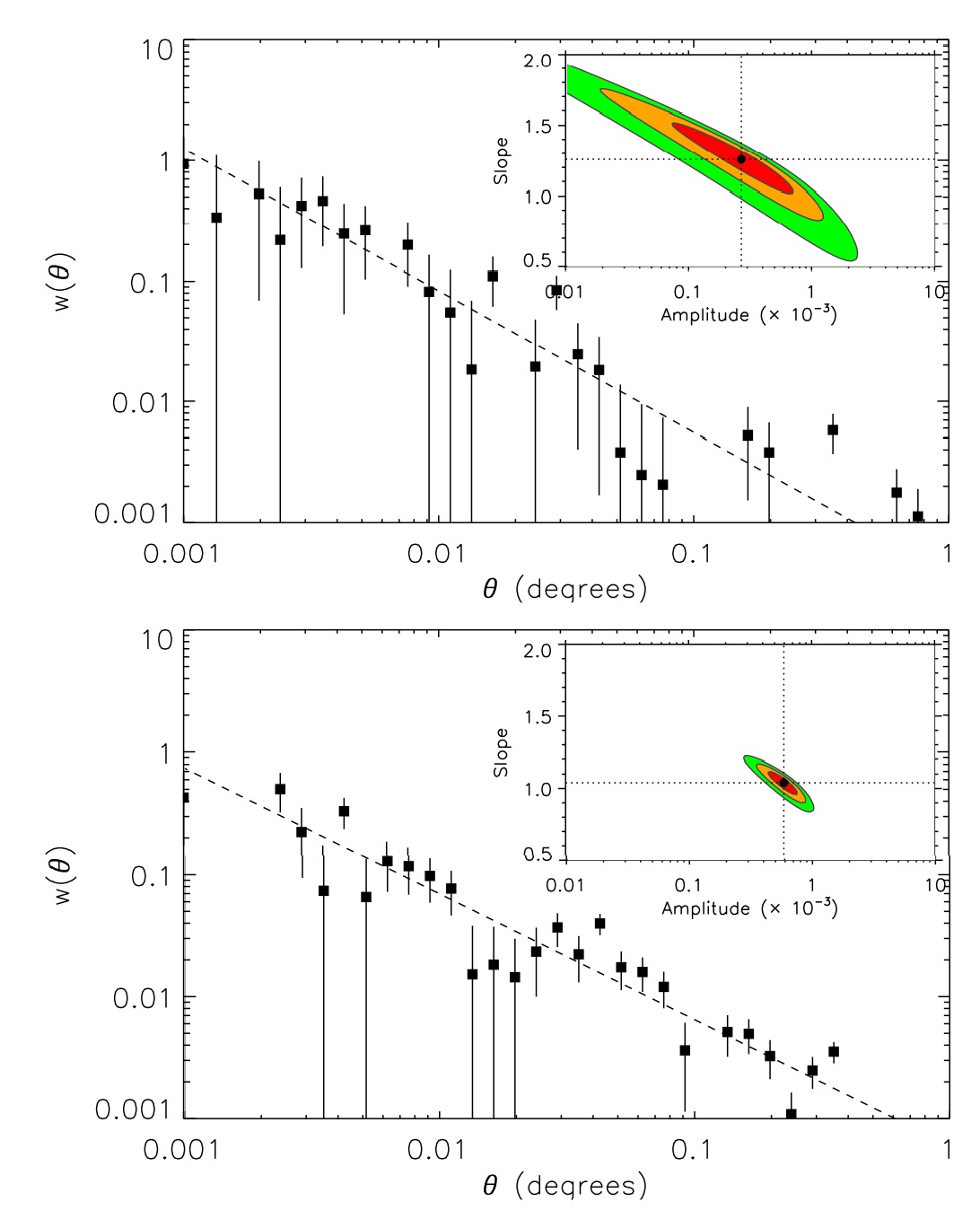}}
\caption{\small\textsf{The angular two point correlation function for a $S>50\mu$Jy radio source (top) 1.5 sq.deg, and (bottom) 4.5 sq.deg. The inset shows the uncertainty contours on the slope and normalization of the power-law fit (Lindsay \& Jarvis in prep.)}}\label{fig:clustering}
\end{wrapfigure}

\noindent
This problem can be tackled in two ways with a deep JVLA continuum survey. The first is to measure the two-point correlation function of the sources in the survey directly. This is analogous to what has been done at low redshifts, where the optical counterparts can be used in these deep fields to determine redshifts, using either photometric or spectroscopic redshifts. Such an experiment requires the necessary volume to determine the clustering of dark matter haloes, and with a single field of around 1 degree$^2$ such a measurement is extremely difficult. However, by moving to 4-5 degree$^2$ patches of sky then the two-halo term in halo-occupation distribution models begins to be measured at $> 1$~Mpc scales. Fig.~\ref{fig:clustering} shows the constraints that can be achieved by moving from a 1.5 degree$^2$ survey to a 4.5~degree$^2$ survey based on the clustering model prescribed in the SKADS simulation \citep{2008MNRAS.388.1335W}. Additional information can also be used, such as the full galaxy catalogue from optical and near-infrared data. By measuring the cross-correlation of the much more abundant optical/near-infrared sources with the radio sources one can obtain much tighter measurements of the clustering of radio sources over all luminosity r\'egimes, i.e. even for the rarer AGN \citep[e.g.][]{2010MNRAS.407.1078D}.

\noindent
A second method to directly trace the density field at these redshifts is attainable using the CMB lensing maps recently determined using data from SZ-telescopes \citep[e.g. the Atacama Cosmology Telescope; ][]{2011PhRvL.107b1301D}. Both the number density of radio sources and the strength of CMB lensing in a certain direction depend on the projected dark matter density in this direction, and conveniently radio sources are most common at the redshifts that produce the largest lensing deflections. This implies that the CMB lensing and radio source fields should be strongly correlated \citep{2000ApJ...540..605P}. Measuring the cross-power spectrum and comparing it to theoretical calculations, would provide the proportionality factor which relates a fluctuation in matter density to a fluctuation in radio source number density, i.e. the bias. 

\noindent
The redshift range over which the CMB lensing signal is most prominent lies at $1 < z < 4$, ideal for cross-correlating with radio sources whose redshift distribution closely follows this lensing kernel. Indeed, a similar study has recently been carried out by the ACT team, where the CMB weak lensing map was cross-correlated with quasars \citep{2012PhRvD..86h3006S}, whose redshift distribution also peaks around $1 < z < 2$. They detect CMB lensing quasar cross-power spectrum
for the first time at a significance of $3.8\sigma$, with a bias of $b = 2.5 \pm 0.6$ (assuming a template for its redshift dependence).

\noindent
Furthermore, the optical and near-infrared counterparts to the radio sources can be used to determine photometric redshifts for the radio sources that enable the redshift distribution of the radio sources to be modified such that the source concentration is shifted in redshift space. This would allow the bias to be determined for the high-redshift ($z > 1$) radio source population, which is expected to be dominated by more highly biased AGN, separately from the low-redshift ($z < 0.5$) source where the source counts become increasingly made-up of less-biased star-forming galaxies. Such an investigation has already been undertaken, based on simulations, by \cite{2012MNRAS.427.2079C} which suggests such a separation provides a unique insight in the very large-scale at high redshift.

\noindent
The case for a weak lensing survey as part of the VLASS \citep[see ][]{BrownVLASS} would also fit naturally with the survey presented here, although this would mean that the deep 10~degree$^2$ survey would need to be carried out in A array in order to at least approach the required resolution for measuring galaxy shapes.

\subsection{The Polarized Sky - Cosmic Magnetic Fields}

Current studies of the deep polarized sky show that it presents questions which cannot be answered by the strong source population alone \citep{2009ApJ...702.1230T}. Indeed \cite{2002A&A...396..463M} found that the mean fractional polarization of radio sources is anti-correlated with flux density. This result was confirmed by \cite{2004MNRAS.349.1267T}, and again by \cite{2007ApJ...666..201T} who found that in a sample of 83 sources with a limiting flux density of $500$~$\mu$Jy the faint source population has a mean fractional polarization almost an order of magnitude larger than for sources $> 100$~mJy. The faint end of this flux density range contains relatively more objects below the FRI/FRII luminosity boundary than the bright end. This raises a number of questions about the nature and evolution of the faint polarized sources. How is the higher degree of polarization of faint sources related to source structure, radio luminosity, redshift, or environment? Does the trend of increasing polarization continue to lower flux densities? When does it stop? Does the polarization of all AGN increase with flux density, or can we identify a sub-class of AGN that is responsible for this trend?
\begin{wrapfigure}{l}{9.5cm}
\resizebox{9.3cm}{!}{\includegraphics{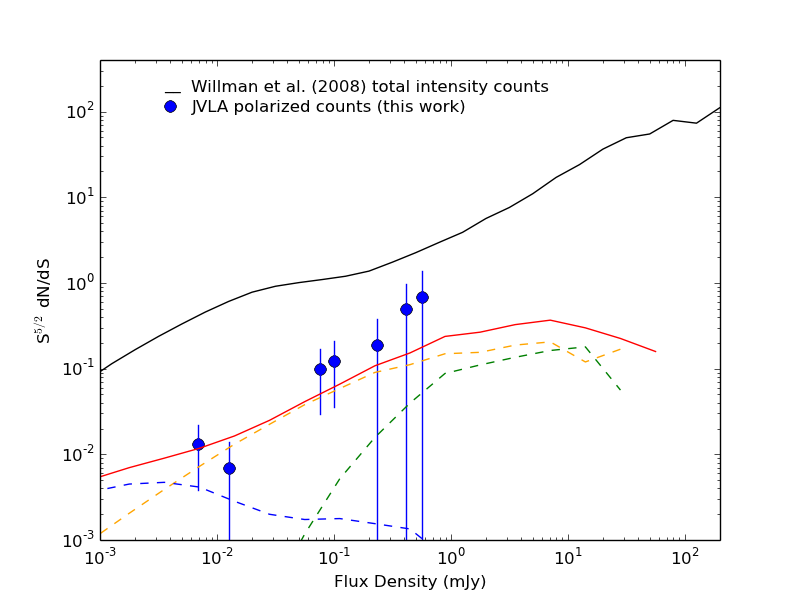}}
\caption{\small\textsf{Polarized intensity source counts over a 0.1 sq deg field at 5 GHz  (Taylor et al.\ in prep.) The solid black line is the \cite{2008MNRAS.388.1335W} SKADS simulation of total intensity counts. 
The red curve is the predicted total polarized flux density counts. The dashed curves are predicted counts  for FR II  and FRI radio galaxies (green and orange respectively), and for polarized emission from normal galaxies (blue).  The proposed survey at 1.5 GHz will directly 
detect thousand of polarized sources spanning the transition where SF and 
normal galaxies become a major component of the polarized population. }}\label{fig:polcount}
\end{wrapfigure}

\noindent
The fractional polarization and intrinsic polarization angle of a source measure the order and direction of its magnetic field. It is expected that these quantities will be different for distinct classes of object due to the changed origin of the emission. In AGN they will mainly be related to the ordered magnetic field in the jets and lobes; for star-forming galaxies, which are mostly spirals, they are likely to reflect the degree of ordering in the intrinsic disk field.
The fractional polarization distribution of nearby disk galaxies at 4.8 GHz was measured by 
\cite{2009ApJ...693.1392S} and \cite{Mitchel2010}, who carried out a survey of nearby (within 100 Mpc)  
but unresolved galaxies using the  Effelsberg Telescope.  These data
show that at at least 60\% of unresolved normal spiral galaxies are polarized 
higher than 1\%, and in some cases higher than 10\%. 
Moreover, there is a strong correlation between polarization position angle and the optical minor axes
of the galaxy disk.
A deep continuum survey over $\sim 10$~square degree with the JVLA would provide direct detection of the presence and properties of ordered  magnetic fields in galaxy disks to intermediate redshifts.

\noindent
The combined areal coverage and depth of the survey, will also allow the polarized source population to be measured to flux-density limits an order of magnitude deeper than existing surveys. Moreover, polarized source counts can me measured to nanoJy levels by stacking the polarized emission at positions corresponding to total intensity detections. In combination with the large optical/IR datasets from VISTA, VST and {\em Herschel} we will be able to cleanly separate star-forming galaxies from AGN in the same way as has been made possible with SDSS in the Northern hemisphere, but with much deeper data at all wavelengths. This multi-wavelength information can then be used to correlate the radio polarization properties with optical emission line diagnostics, galaxy type and star-formation rates. This is important as there is already evidence that magnetic fields may be important in controlling the global star formation properties of spiral galaxies.

\noindent
Such a large population of polarized sources will allow polarization properties of galaxies to be used for cosmological applications. One of the long term aims of polarization surveys is to investigate the alignment between large-scale structure and the polarization position angle of the polarized source population. Preliminary investigation using the NVSS survey, which is dominated by jet-powered AGN, finds that there are significant correlations between the position angle of the total intensity isophotes and the polarization position angle. The proposed survey will allow us to probe substantially lower flux densities at higher resolution. Importantly, this will allow us to remove the effects of internal Faraday rotation, which strongly suppresses the intrinsic polarization and makes the correlation detectable only by stacking a substantial number of galaxies.

\noindent
In addition it should also be possible to use the data to put stringent limits on the rotation of the plane of polarization due to the coupling of the electromagnetic sector to pseudo-Nambu Goldstone bosons, so called cosmic birefringence \citep{1998PhRvL..81.3067C}.
By performing rotation measure (RM) synthesis, or otherwise extracting rotation measures for the sources, it will also be possible to statistically examine the distribution of RMs to search for the imprint of intergalactic fields on galaxies on cosmological scales. We will be sensitive to the internal Faraday rotation of the sources. The amount of depolarization by Faraday dispersion sets apart star-forming galaxies from most AGN powered radio sources potentially allowing us to separate them using the polarization observations; something which can be cross-checked using the multi-wavelength information.  

The bandwidth of the survey will allow Faraday Rotation Measures to be measured
with an accuracy better than 2 rad-m$^{-2}$ down to polarized flux densities of 20 $\mu$Jy. 
From recent JVLA imaging to 1 $\mu$Jy rms of a 0.1 sq deg field at 5 GHz (Figure~\ref{fig:polcount}),
we expect to detect several 100 sources per square degree down to 10 $\mu$Jy at 1.5 GHz. 
The 2-point correlation function of RM will be measured on scales of arc minutes and precision of 
a few  rad-m$^{-2}$ , precisely the regime where fluctuations from primordial magnetic fields are 
expected to create detectable signal \citep{1998ApJ...495..564K,2010ApJ...723..476A}.  This regime of sky
density and RM accuracy has not been accessible to date.

\subsection{H{\sc i} Deep Field}

By conducting the survey at L-band rather than at S-band, we not only obtain a higher source density (approximately a factor of 2.5 higher at L-band than for S-band given the same observing time), crucial to the key science aims set out above, but also enable a complementary H{\sc i} survey at the same time. Although we do not present the case for the H{\sc i} survey here we emphasise that the science enabled by conducting such a survey over the best studied deep fields holds for H{\sc i} as it also does for continuum. Any H{\sc i} deep field survey would therefore likely choose the same fields as we do here.

\noindent
It would also be possible to undertake a deep search for H{\sc i} absorption against moderately bright radio sources within our fields, along with a coarse velocity resolution search for other lines, such as OH megamasers.  A search for H{\sc i} absorption could be carried out  for sources at $z<0.5$.  These H{\sc i} absorption studies are not limited by the brightness sensitivity to H{\sc i} emission (typical T$_{\rm spin}\,\approx$\,100\,K) but rather the spectral line sensitivity to detect absorption signals against high T$_{\rm spin}$ continuum sources. Thus it is possible to detect H{\sc i} via absorption at much higher redshifts and with higher angular resolutions than is possible for typical H{\sc i} emission experiments.

\subsection{Transients}

\noindent
The continuum survey would be structured in such a way that it will be possible to find very slow and rapid transient phenomena. The many repeat observations will allow AGN to be found via variability to study the time domain astrophysics of extragalactic sources. Such a science case would fit naturally within a VLASS deep continuum survey.

\section{Field Selection:  Multi-wavelength Synergies and Long-term Legacy Value}\label{sec:multilambda}

\noindent
The survey fields have been carefully chosen to ensure maximum overlap over fields which have unrivalled multi-wavelength data sets and that are extremely important in fulfilling the continuum survey science aims.
The crucial data set for the continuum survey is information on the redshift of the extragalactic radio sources. Redshifts can be obtained either by spectroscopy or using photometric redshifts based on multi-band optical and/or near-infrared data.

\noindent
We suggest observing three fields which will continue to be the most widely observed fields in extragalactic astronomy. The COSMOS, XMMLSS, ECDF-S have all been observed by {\em Spitzer} as part of the SWIRE survey \citep{2003PASP..115..897L}, with the data fully reduced and available. They have also being observed as part of the HerMES Guaranteed Time Programme on {\em Herschel} \citep{2012MNRAS.424.1614O} which supplies deep imaging data from 100--500$\mu$m. The deep radio data from the VLASS will be crucial for fully interpreting the {\em Herschel} data set due to the much higher resolution which will allow us to associate the {\em Herschel} sources with the correct optical and near-infrared identifications. These fields will also be covered at near-infrared wavelengths by the Ultra-VISTA Survey \citep{2012A&A...544A.156M} over the COSMOS field \citep{Scoville2007}, and the VIDEO Survey \citep{2013MNRAS.428.1281J}. These fields have also been surveyed by the {\em Spitzer} Representative Volume Survey \citep[SERVs; ][]{2012PASP..124..714M} and at visible wavelengths they are part of the Dark Energy Survey (DES) Supernova Survey \citep{Bernstein2012}, and a VST Deep Survey as part of Guaranteed time.
They will also form part of the LSST Deep Drilling Fields.
As such these fields will have exquisite optical data tand near-infrared data, providing at least 20-band photometry for measuring photometric redshifts out to $z \sim 5$ for typical elliptical galaxies and beyond $z \sim 6$ for the most massive galaxies which host radio-loud AGN.
In terms of more specific science aims all of these fields also have X-ray data from both {\em XMM-Newton} and {\em Chandra} allowing AGN to be found by virtue of their X-ray emission, {\em GALEX} data which provides UV-photometry.

\begin{table*}
\noindent\caption{\small\textsf{Multi-wavelength data available over the key extragalactic fields accessible to the deep VLASS continuum survey. All of the optical and near-infrared data in these fields are deep enough to estimate photometric redshifts for upward of 70 per cent of the radio sources at a flux-density threshold of 7.5$\mu$Jy at 5$\sigma$; \citep[see e.g.][]{2012MNRAS.423..132M}. All fields are also covered by the {\em Herschel} with the HerMES survey.}}
\begin{tabular}[h]{p{20mm}p{30mm}p{8mm}p{25mm}p{25mm}p{20mm}p{25mm}}
\hline\hline
Field & Coordinates & Area & Optical Imaging & Near-IR & Mid-IR &  Spectroscopy \\
\hline
COSMOS & 10 00 29 +02 12 21 & 1.5 & Subaru, CFHT, DES & UltraVISTA & SCOSMOS   & zCOSMOS \\
XMMLSS & 02 22 00 -04 48 00 & 4.5 & Subaru, CFHT, DES & VIDEO, UKIDSS-UDS & SWIRE, SERVS & VVDS, OzDES \\
ECDFS & 03 30 08 -28 38 00 & 4.5 & DES, VST & VIDEO & SWIRE, SERVS & OzDES \\
\hline\hline
\end{tabular}
\end{table*}

\section{Observational Strategy}\label{sec:strategy}

\begin{figure*}
\begin{center}
\resizebox{15.cm}{!}{\includegraphics{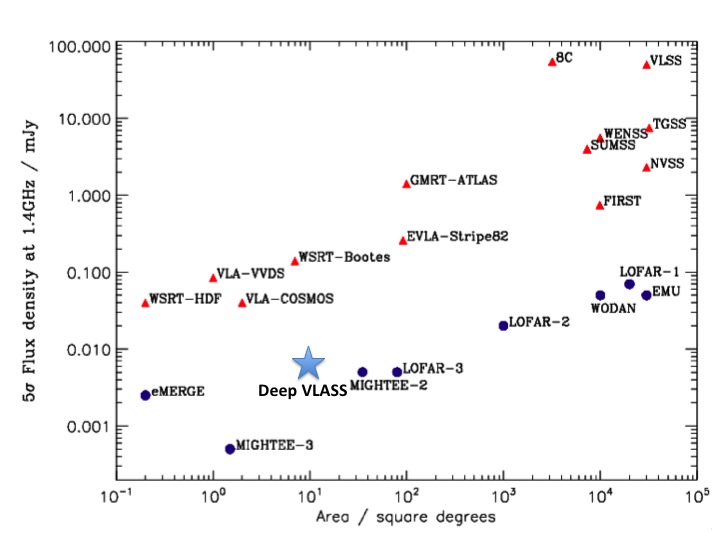}}
\caption{\small\textsf{Current and proposed radio surveys on existing facilities and those proposed for the SKA Precursor telescopes. The JVLA Deep Survey that is detailed here is denoted by the star.}}
\end{center}
\end{figure*}

The VLASS continuum survey is most efficiently conducted at L-band, even with the effective bandwidth of 600~MHz compared to the 1.75~GHz of bandwidth available in S-band. This is due to a combination of enhanced field of view at L-band combined with the typical spectral index for extragalactic radio sources of $\alpha = -0.7$ (where $S_{\nu} \propto \nu^{\alpha}$). For example, for an rms flux-density limit of 1.5$\mu$Jy at L-band, we would need to reach $\sim 0.9\mu$Jy at S-band to sample a similar source population density. Although this depth can be reached in roughly half the time than for a survey to 1.5$\mu$Jy at L-band, the primary-beam area is reduced by a factor of four. Thus the survey speed is essentially a factor of two higher at L-band compared to S-band for a survey covering $>1$ primary beam element.  

\noindent
As stated in Section~1.6 this also means that both H{\sc i} emission and absorption surveys can be conducted at the same time, thus enhancing the scientific value of the survey, without effecting the key continuum and polarization science emphasized here.

\noindent
However, we note that some science aims push the case for high-resolution in order to carry out a weak lensing survey and also to enable the separation of star-formation and AGN activity in galaxies through morphological measurements. Thus there are obvious trade-offs in the frequency and array configuration in order to meet these goals. If the H{\sc i} survey component is deemed unnecessary or that the WIDAR correlator could not perform the spectral line and continuum survey at the same time then a case for S-band would be much stronger given the enhanced resolution which makes it very distinct from the SKA precursors, which will have a resolution of 5-10~arcsec and thus not be able to categorize high-redshift sources on the basis of their radio morphology.

\noindent
To cover 10~degree$^2$ requires the equivalent of 70 pointings. To reach the 1.5$\mu$Jy rms sensitivity with 600~MHz of effective bandwidth requires 57 hours per pointing. Thus the total survey would require $71 \times 57 = 4047$ hours. However, we note that the COSMOS field is part of the CHILES survey, and also other fields already have deep JVLA data to around 5-10$\mu$Jy. Thus the total time request will be of the order 4000~hours (including overheads).

\noindent
In Fig.~5 we show how the suggested survey fits in with current and planned radio continuum surveys in terms of the depth versus area. In terms of current or planned deep continuum survey, one can immediately see that the VLASS deep continuum survey is very competitive with the planned deep continuum survey \citep[MIGHTEE; ][]{Jarvis2012} for the MeerKAT telescope. Indeed, it will reach a greater depth than Tier-2 of the MIGHTEE survey, albeit with a factor of $\sim 3$ less area. Given the current baseline distribution for the MeerKAT telescope, Tier-3 of MIGHTEE will not be feasible as it will become confusion limited before it reaches the final depth. The ASKAP-EMU survey \citep{2011PASA...28..215N} is concerned with a different part of parameter space and is focussed on wide-field ($\gg1000$ of square degrees) science. Thus, the VLASS deep continuum survey would fill a unique part of parameter space well before MeerKAT is in operation and well before such a survey could be taken to a new level by the SKA itself.

\subsection{Dynamic Range}

\noindent
Achieving a noise-limited, full-Stokes mosaic image in the presence of the numerous sources that will be a good fraction of a Jy expected in the survey area, will require application of frequency and direction dependent calibration over the primary beam of the JVLA,
including the accurate treatment of sources in the sidelobes. Future radio surveys will routinely reach depths where deficiencies in the traditional approach to calibration are the limiting factor, and overcoming this is an active and productive area of research (for several co-authors of this white paper).

\begin{figure*}
\begin{center}
\resizebox{15.cm}{!}{\includegraphics{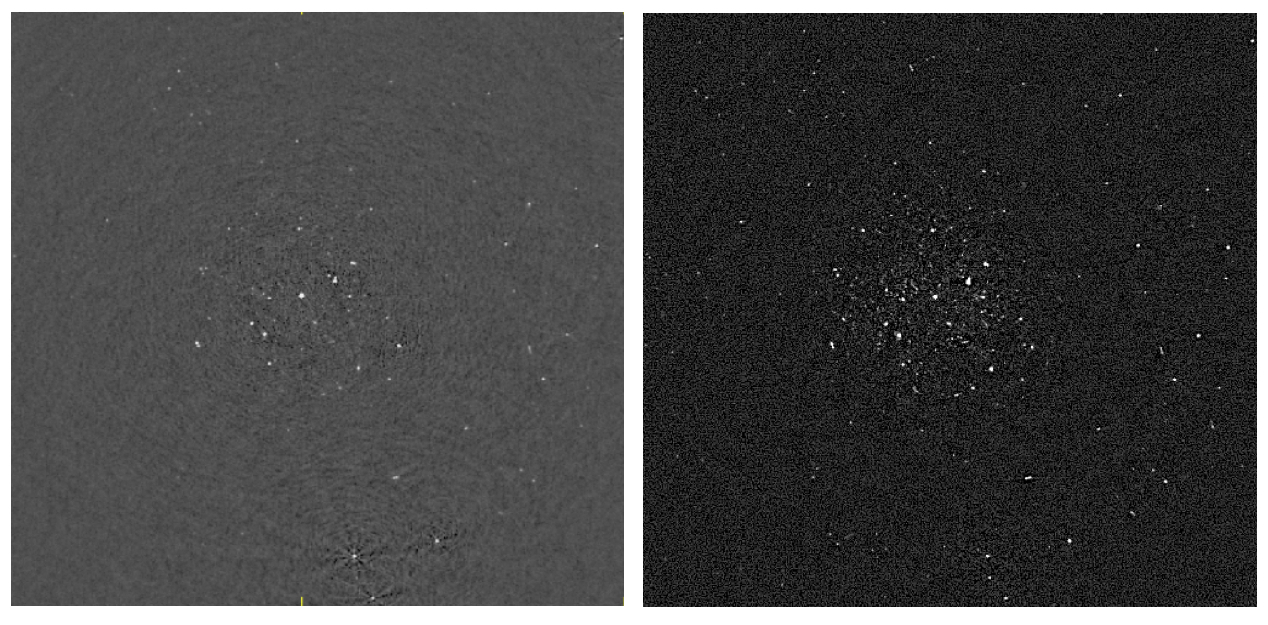}}
\caption{\small\textsf{The 3C147 field imaged using VLA C+D array data using traditional self calibration methods (left) and direction dependent calibration using the MeqTrees software system (right). Note that part of the increased depth in the latter image comes from using three sub-bands instead of one, but the salient point here is that the PSF-like artefacts associated with off-axis sources (particularly in the first sidelobe) are completely eliminated. The map is thermal noise limited throughout, with a dynamic range of 3,200,000:1 \citep{Perley2013}.}\label{fig:3C147}}
\end{center}
\end{figure*}

\noindent
The need for direction dependent calibration methods \citep{2010A&A...524A..61N,2011A&A...527A.108S} for pathological scenarios where traditional direction-independent calibration methods are inadequate has already been demonstrated. One such example is that of extreme dynamic range imaging (Perley \& Smirnov, 2013, see also Figure \ref{fig:3C147}), another involves the modelling and subtraction of a phase calibrator in the first null of the primary beam, and in the presence of pointing errors \citep{2013MNRAS.428..935H}. We are currently developing pipelines to automatically apply these methods to moderately deep (XMM-LSS, 10~$\mu$Jy, 4.5~deg$^{2}$) and wide (Stripe 82, 50~$\mu$Jy, 100~deg$^{2}$) VLA surveys in order to mitigate the effects of strong sources in the field (Heywood, Jarvis, Smirnov, et al., in prep).

\noindent
The derived gain solutions make physical sense, showing conclusively that time and frequency dependent beam gain variations are the dominant direction dependent effect even at L-band. Noise-limited Stokes-I images produced using these techniques can offer increased scientific return, however deficiencies remain for Q, U and V imaging. Efforts to address this are focused on deriving initial (and refined) sky models in full polarization using A-projection imaging within the calibration cycle. A-projection applies prior models of the beam patterns during gridding and has been shown to vastly improve polarization performance, correcting for frequency dependent off-axis complex gain and polarization leakage as a function of parallactic angle \citep{2008A&A...487..419B}.

\noindent
This work will require the survey team working with the advanced algorithm research and development (ARD) group on the implementation and commissioning of broad-band A-projection corrections within CASA.  These algorithms will  correct for frequency dependent off-axis complex gain and polarization leakage as a function of parallactic angle.   They can then be incorporated into a CASA-based processing pipeline for broad-band, full Stokes mosaic imaging developed as part of an EVLA Resident Shared Risk observing program.

\noindent
Combined A-projection, with frequency dependent corrections over broad bands and joint deconvolution of multiple mosaic pointing is a computationally intensive problem.  Therefore part of the work with the ARD group could include implementation of the CASA pipeline into HPC and/or massively parallel computing architectures.

\section{Outreach and Citizen Science}

There has recently been a large amount of effort devoted to constructing ``Radio Galaxy Zoo'', to be launched over the 2013-2014 holiday period. The aim of this Citizen Science project is to determine the morphology and cross-identification of radio sources from a range of radio surveys. These data are currently coming from wide-area surveys and being overlaid on optical imaging from the SDSS for example. With the VLASS deep continuum surveys, such a project could be taken to the high-redshift Universe. The 10 square degree scale surveys along with the VLASS continuum survey could be incorporated into Radio Galaxy Zoo and allow the public to contribute to high-redshift science in enabling cross-matching and characterisation of many 10s of thousands of radio sources at $z>1$.

\bibliographystyle{mn2e}
\small{
\bibliography{VLASS_deep_final}
}
\newpage

\noindent
$^{1}$Astrophysics, University of Oxford, Denys Wilkinson Building, Keble Road, Oxford, OX1 3RH, UK\\ 
$^{2}$Department of Physics, University of the Western Cape, Bellville 7535, South Africa\\ 
$^{3}$National Radio Astronomy Observatory, Socorro, NM, 87801, USA\\
$^{4}$Hamburger Sternwarte, Universit\"at Hamburg, Gojenbergsweg 112, D-21029 Hamburg, Germany\\
$^{5}$Laboratoire Lagrange, UMR7293, Universit\'e de Nice Sophia-Antipolis, CNRS, Observatoire de la C\^ote d'Azur\\
$^{6}$CSIRO Astronomy \& Space Science, P.O. Box 76, Epping, NSW 1710, Australia\\
$^{7}$Centre for Radio Astronomy Techniques and Technologies (RATT), Department of Physics and Electronics, Rhodes University, PO Box 94, Grahamstown 6140, South Africa\\
$^{8}$School of Physics, Astronomy and Mathematics, University of Hertfordshire, College Lane, Hatfield AL10 9AB, UK\\
$^{9}$Infrared Processing and Analysis Center, California Institute of Technology, MC 220-6, Pasadena, CA 91125, USA\\
$^{10}$Astrophysics, Cosmology Gravity Centre, University of Cape Town, Cape Town, 7701, South Africa\\
$^{11}$Department of Physics and Astronomy, University of Calgary, 2500 University Drive NW, Calgary, Alberta, T2N 1N4, Canada\\
$^{12}$SKA South Africa, 3rd Floor, The Park, Park Road, Pinelands 7405, South Africa\\
$^{13}$Astrophysics Research Institute, Liverpool John Moores University, Twelve Quays House, Egerton Wharf, Birkenhead CH41 1LD, UK\\
$^{14}$University of Zagreb, Physics Department, Bijenicka cesta 32, 10002 Zagreb, Croatia\\

\end{document}